\documentstyle[aps,preprint]{revtex}
\tightenlines

\begin{document}
\draft
\title{An Interpretation of the Neutron Scattering Data on Flux Lattices of
Superconductors}
\date{\today }
\author{D. Chang$^*$, C.-Y. Mou$^*$, B. Rosenstein$^\dagger$, and C. L. Wu$^*$}
\address{$^*$ Department of Physics,
National Tsing Hua University,
Hsinchu, Taiwan, 30043, R.O.C. \\
$^\dagger$ Electrophysics Department,
National Chiao Tung University,
Hsinchu, Taiwan, 30043, R.O.C.}
\maketitle

\begin{abstract}
Small angle neutron diffraction experiments are analyzed using recently
developed and properly generalized one-field effective free energy method.
In the case of experiment of Keimer et al on YBCO, we show that the fourfold
symmetry of the underlying crystal is explicitly broken, but the reflection
with respect to the [110] and [1\={1}0] axes remains a symmetry. The vortex
lattice also becomes generally oblique instead of rectangular body centered.
Unexpectedly rich phase diagram is described.
\end{abstract}


There are growing evidences that superconductivity in layered high $T_c$
cuprates is largely due to the $d_{(x^2-y^2)}$ pairing\cite{Annett} with
small mixing of s-wave component\cite{Dagotto,Tsuei,Martindale}. The
unconventional pairing mechanism makes an impact on the single vortex and
the vortex lattice structure. Recent studies on the detailed structure of
the Abrikosov vortex lattice in YBCO, using small angle neutron diffraction 
\cite{Yethiraj,Keimer} and tunneling spectroscopy \cite{Maggio}, show clear
deviations from the standard triangular lattice. It is natural to try to
explain these deviations theoretically with modified phenomenological
Ginzburg - Landau (GL) theory. To investigate d-wave superconductors with
s-wave mixing, Ren et al. \cite{Ting1} and Soininen et al. \cite{Berlinsky1}
both derived an effective GL type theory using two order parameters: $s$ and 
$d.$ From this effective action, or more fundamental equations\cite{Ichioka}%
, one obtains a characteristic four-lobe structure for an isolated vortex
and its associated magnetic field\cite{Berlinsky2}. The fourfold vortex core
structure comes into conflict with the high symmetry of the triangular
lattice and can distort it at already accessible fields much lower than $%
H_{c2}$. The vortex lattices obtained within this approach are basically
centered rectangular lattice with chains of vortices oriented along
crystalline axes [100] and [010] (see Fig. 1). They spontaneously break the
fourfold rotational symmetry (i.e., two different lattices related by $90^o$
rotation), but preserve the reflections with respect to the axes [100] and
[010].

These predictions come close to results of some experiments \cite
{Yethiraj,Maggio}, but clearly disagree with those of \cite{Keimer}.
According to the interpretation given in \cite{Keimer}, the centered
rectangular vortex lattice gets rotated by $45^{\circ }$ with respect to the
crystalline axes (see Fig. 2a), i.e., the chains of vortices lie along the
diagonal directions [110] and [1\={1}0] instead. A recent theoretical study
by Ren, et al.\cite{Ting3} has considered explicit breaking of the fourfold
symmetry within the two field framework. Their results however remains
qualitatively the same as the case with fourfold symmetry - only centered
rectangular nonrotated vortex lattices are obtained. So far, there is no
theoretical interpretation for the lattice data observed in \cite{Keimer}.
We shall provide such an interpretation in this letter. Our answer is
different from that provided in \cite{Keimer}, however, the results can
still be derived from the GL theory with proper fourfold symmetry breaking
terms.

In this work we adopt a recently developed one field effective theory, first
introduced by Affleck et al \cite{Affleck} in which they work mainly in the
London limit, and later by us \cite{Chang} for static and moving vortex
lattices near $H_{c2}$. Most of the above mentioned results can be
reproduced in this much simpler formulation in which only the field $d$ is
introduced and the theory is based on the following $D_{4h}$ symmetric free
energy 
\begin{equation}
F_{eff}[d]=\frac 1{2m_d}|\Pi d|^2-\alpha _d|d|^2+\beta |d|^4-\eta
d^{*}\left( \Pi _y^2-\Pi _x^2\right) ^2d,  \label{fe}
\end{equation}
where $\Pi =-i\nabla -e^{*}{\bf A}$. The last term which we call $F_{4d}$
parametrizes the breaking of full rotational symmetry down to $D_{4h}$ and
can be treated as a perturbation. Near $H_{c2}$, the linearized equation in
the one field approach can be solved perturbatively in $\eta $, which allows
one to easily generalize the description of the centered rectangular
lattices to the most general oblique lattices\cite{Chang}. This will be
crucial in the present work in which these more general lattices are indeed
the ground state in some cases. 

Note that the contributions to the coefficient $\eta $ might not only come
from the d-s mixing which always gives a positive $\eta $, but also from
other sources \cite{Chang}. The possibility of having negative $\eta $ will
be discussed later. It is also important to realize that since this
formulation utilizes only the symmetry properties, it can be applied to the
conventional type II superconductors with $D_{4h}$ symmetry as well. In this
case,  $\eta $ is proportional to the angular average of products of Fermi
velocities on the Fermi surface, describing the deviation of the Fermi
surface from a perfect sphere\cite{Hohenberg}. The effective strength of $%
F_{4d}$ can be characterized by a dimensionless parameter $\eta ^{\prime
}\equiv \eta m_de^{*}H$\cite{Chang}. Using the free energy in Eq.(\ref{fe}),
one finds centered rectangular vortex lattices (see Fig. 1) with the angle $%
\alpha $ directly related to the coefficient $\eta ^{\prime }$. The lattice
becomes square when $\eta ^{\prime }$ exceeds a critical value $\eta
_c^{\prime }=.0235$\cite{Chang}. This can accommodate the tunneling
spectroscopy data of \cite{Maggio} and the SANS data of \cite{Yethiraj} for
YBCO, as well as a recent decoration and neutron scattering data for a low $%
T_c$ material ErNi$_2$B$_2$C\cite{Eskildsen}. The analysis presented in \cite
{Chang} indicates that the precise SANS data of \cite{Eskildsen}
unambiguously shows that for large $\eta ^{\prime }$, the vortex lattice
becomes a square one, exhibiting perfect $D_{4h}$ symmetry. The less precise
data of \cite{Maggio} gives an angle $\alpha \approx $ $54^o$, which
corresponds to $\eta ^{\prime }\approx 0.019$ and is in the centered
rectangular phase. These two experiments on two different samples both seem
to show manifestation of $D_{4h}$ symmetric GL free energy and correspond to
its two different phases. The transition from the centered rectangular
vortex lattice to the square lattice  was observed in $ErNi_2B_2C$\cite
{Eskildsen} and has not been observed yet in high $T_c$ materials.

In order to explain the data in \cite{Keimer}, we now generalize the
formalism to include terms which break the $D_{4h}$ symmetry. This can be
also motivated by noting that in many high $T_c$ cuprates the $D_{4h}$
symmetry is not exact. For example, the CuO chains in YBCO\ breaks the
fourfold symmetry down to twofold\cite{Ting3}. Up to (scaling) dimension
three, there are two possible terms that break fourfold symmetry: $%
F_{x^2-y^2}=-\mu d^{*}(\Pi _y^2-\Pi _x^2)d$ and $F_{xy}=-\lambda d^{*}(\Pi
_x\Pi _y+\Pi _y\Pi _x)d.$ The first term $F_{x^2-y^2}$ describes the
asymmetry between [100] and [010] axes and has the reflection symmetries $%
x\rightarrow -x,y\rightarrow y$ ($\sigma _x$) and $x\rightarrow
x,y\rightarrow -y$ ($\sigma _y$). This term has already been considered in 
\cite{Ting3}. The second term $F_{xy}$, on the other hand, preserves the
reflection symmetry with respect to the [110] and [1$\overline{\text{1}}$0]
directions, that is, $x\rightarrow y,y\rightarrow x$ and $x\rightarrow
-y,y\rightarrow -x$. In the BCS theory, the presence of the second term
requires that the shape of the Fermi surface also breaks the $\sigma _x$ and 
$\sigma _y$ symmetries . Since this is quite unlikely, we do not expect that
it will occur in the conventional superconductors. We will find, however, in
the case of Keimer et al's SANS experiment, the $F_{xy}$ term is required to
explain the data.

The method of calculation is quite analogous to that of the $\eta $
correction explained in \cite{Chang}, so here we just present the result.
Let $a,b$ be the two lattice constants and $\alpha $ be the angle between
the two basis vectors (Fig.1). It will be convenient to introduce the
complex variable $\zeta \equiv \frac{b}{a}e^{i\alpha }\equiv \rho +i\sigma $%
. The angle between the vortex lattice and the crystalline lattice will be
denoted by $\varphi $. The Abrikosov's $\beta _{A}\equiv \left\langle \left|
d\right| ^{4}\right\rangle /\left\langle \left| d\right| ^{2}\right\rangle
^{2}$ is then given by: 
\begin{eqnarray}
\beta _{A}\left( \rho ,\sigma \right) &=&\beta _{A}^{0}\left( \rho ,\sigma
\right) +\frac{\sqrt{\sigma }}{4}%
\mathop{\rm Re}
\left\{ \left[ \sum_{n^{\prime }=-\infty }^{\infty }\exp (-2\pi i\zeta
^{*}n^{\prime 2})\right] \left[ \sum_{n=-\infty }^{\infty }\exp (2\pi i\zeta
n^{2})G(n)\right] \right. +  \nonumber \\
&&\left. \left( n\rightarrow n+\frac{1}{2},n^{\prime }\rightarrow n^{\prime
}+\frac{1}{2}\right) \right\} .
\end{eqnarray}
where $\beta _{A}^{0}\left( \rho ,\sigma \right) $ can be found in standard
textbook or in \cite{Chang}. All the three anisotropic corrections are
collected in the prefactor: 
\begin{eqnarray}
G(n) &=&\eta ^{\prime }e^{4i\varphi }(64\pi ^{2}\sigma ^{2}n^{4}-48\pi
\sigma n^{2}+3)  \nonumber \\
&&+4\mu ^{\prime }e^{2i\varphi }(8\pi \sigma n^{2}-1)  \nonumber \\
&&+4\lambda ^{\prime }e^{2i\left( \varphi +\pi /4\right) }(8\pi \sigma
n^{2}-1)  \label{G_n}
\end{eqnarray}
where $\mu ^{\prime }\equiv \mu m_{d},\lambda ^{\prime }=\lambda m_{d}$.

The term $F_{y^{2}-x^{2}}$ in the effective energy preserves the symmetries
of the centered rectangular lattice and is therefore not expected to produce
interesting qualitative effects, so we will drop the $\mu ^{\prime }$ term
in the following discussions, however it is understood that in making
quantitative comparison with data, the $\mu ^{\prime }$ term may have to be
included. The remaining correction to $\beta _{A}$ summarized in $G(n)$ has
two parts: the first one comes from the fourfold symmetric term $F_{4d}$ and
has $e^{4i\varphi }$ angular dependence. The second one has $e^{2i\left(
\varphi +\pi /4\right) }$ angular dependence and comes from $F_{xy}$. It is
this conflict between the two contributions that gives rise to the observed
diffraction pattern. Either $F_{4d}$ or $F_{xy}$ alone will give reflection
invariant lattices, i.e., rectangular body centered lattice aligned along
[100] or [110], respectively. The lattice structure is determined by
minimizing $\beta _{A}$ with respect to $\rho ,\sigma $ and $\varphi $
numerically. One obtains generally nonrectangular oblique vortex lattices.
It differs markedly from the $D_{4h}$ symmetric case.

Fig. 2(b) shows the diffraction pattern and the corresponding lattice
structure that we obtained at $\eta ^{\prime }=0.019,\lambda ^{\prime }=0.04$%
. (Note that in 2D the reciprocal lattice is nothing but a rotation of $90^o$
of the real lattice). In one of the diffraction patterns, Fig.2(a), one sees
clearly two large peaks in the [110] direction and four weaker points on
both sides of the [110] line, giving totally 10 points. This was interpreted
in Ref.\cite{Keimer} as a nearly rectangular lattice with one of the basis
vector lying on [110], together with its reflected version. (Presumably, the
two lattice orientations are degenerate ground state and show up
simultaneously as different domain in the sample). The points on the [110]
line then coincide and produce constructive interference. In comparison, in
the previous calculations, because the reflection symmetries $\sigma _x$ and 
$\sigma _y$ are preserved, one always obtained rectangular lattices which
are aligned to either [100] or [010]. They possess twofold symmetry and upon
reflection one is unable to produce different lattices. As a result, there
will be only 6 points on the diffraction pattern and can not account for the
data.

We notice an important difference here: The off diagonal points (the four
weaker points) are not really on the line parallel to [110] as Keimer et al.
claimed. One might hope to tune the parameters such that when the two points
on [110] merge into one, the off diagonal points will align themselves as
well, but this is not the case. In fact, they will also merge with each
other, and there will be no splitting anymore. If one look carefully at
their contour plot it is possible to tell the difference. Furthermore, the
lattice we obtained is not rectangular, this is consistent with their
possibly $5\%$ difference between the length of the two primitive basis
vectors.

The vortex lattice phase diagram in the $(\eta ^{\prime },\lambda ^{\prime
}) $ plane is presented in Fig. 3. Since changing the sign of $\lambda
^{\prime }$ only reverses the roles of [110] and [1\={1}0] axes, it suffices
to show only the positive $\lambda ^{\prime }$. First, consider the $D_{4h}$
symmetric case with $\lambda ^{\prime }=0$. Then $\eta ^{\prime }=0$
corresponds to the conventional triangular lattice with no special
orientations. For $\eta ^{\prime }<\eta _{c}^{\prime }$ the lattice is
centered rectangular aligned to [100] and [010] with double degeneracy
(related by reflection about [110]). Increasing $\eta ^{\prime }$\ elongates
lattices along either [100] or [010] so that when $\eta ^{\prime }>\eta
_{c}^{\prime }$, the two degenerate lattices both becomes square and the
full $D_{4h}$\ symmetry is restored.

For $\lambda ^{\prime }>0$, there are three phases and two phase transition
lines. The lattice, compared to the corresponding $\lambda ^{\prime }=0$\
case, can in general be thought of as resulting from a deformation in the
[110] direction. (For $\lambda ^{\prime }<0$, the corresponding deformation
will be in the [1\={1}0] direction). The lattice is centered rectangular for
smaller values of $\eta ^{\prime }$ , while it is rectangular (not centered)
for larger values of $\eta ^{\prime }$. Symmetry of the unique ground state
in each of these two cases is larger than that of the free energy. There is,
however, no direct phase transition between them. Instead, in the region
between these two phases bounded by $\lambda _1^{\prime }(\eta ^{\prime })$
and $\lambda _2^{\prime }(\eta ^{\prime })$, there is a less symmetric phase
in which ground states are doubly degenerate. This comes from a nontrivial
competition between $F_{4d}$ and $F_{xy}$. The two degenerate lattices are
also related by the reflection about [110] and are generally oblique. We see
that the data in \cite{Keimer} can be fitted into this phase. The transition
line $\lambda _1^{\prime }(\eta ^{\prime })$ starts from the origin and
monotonically increases with $\eta ^{\prime }$, while $\lambda _2^{\prime
}(\eta ^{\prime })$ starts from $(\eta _c^{\prime },0)$ and also increases
monotonically. $\lambda _2^{\prime }(\eta ^{\prime })$ appears to approach $%
\lambda _1^{\prime }(\eta ^{\prime })$ asymptotically. Since $\eta ^{\prime }
$ is proportional to the magnetic field $H$, one immediate implication of
this phase diagram is that, for a given sample, by increasing $H$ one should
encounter two phase transitions. This prediction can be tested directly by a
number of experimental techniques.

We would like to briefly describe here another, rather exotic possibility.
The one field approach allows one to consider the negative $\eta $ case.
This cannot be obtained from the two field formulation in which $\eta $ is
always positive if we only assumes one critical temperature\cite
{Ting1,Berlinsky1}. However the possibility of negative $\eta $ cannot be
ruled out theoretically. In the one component theory with exact fourfold
symmetry, the negative $\eta $\ is equivalent to the $d_{xy}$\ pairing,
while in the BCS theory, it could happen if the Fermi surface is elongated
along $y=\pm x$ direction. \ When $\eta ^{\prime }$ is negative, the minus
sign replaces $\varphi $ in the $\exp \left( 4i\varphi \right) $ factor in
Eq. (\ref{G_n}) by $\varphi \pm 45^o$, and then both $F_{4d}$ and $F_{xy}$
will prefer the diagonal direction. As a result there will be no competition
and we will always get rectangular body centered lattices along [110] or
[1\={1}0].

In conclusion, we have investigated the effects of explicit fourfold
symmetry breaking on the Abrikosov lattice structure with the one field
formulation . The complete phase diagram was constructed. We found quite
rich phase diagram with three different phases separated by two phase
transition lines. The vortex lattice observed in Keimer et al.'s experiment%
\cite{Keimer} can be accommodated in the new phase diagram .{\bf \ }It turns
out that the vortex lattices are no longer centered rectangular, but rather
general oblique ones. Other experiments fit quite well into the $D_{4h}$
symmetric phase in which the triangular to square phase transition takes
place.

Authors are very grateful to C.C. Chi, M.K. Wu, W. Yang, and S.Y. Hsu. This
work is supported by NSC of R.O.C. under Grants No. NSC 86-2112-M-007-006
and 86-2112-M-009-034-T.

\clearpage

\noindent {\bf Figure Caption}: \newline
Fig. 1 The body centered rectangular lattice obtained in the fourfold
symmetric case, the two lattices (a) and (b) are related by a rotation of $%
90^{o}$ or reflection about the [110] axis. \newline
\newline
Fig. 2 Keimer et al's SANS diffraction pattern and two different
interpretations. (a) Keimer et al's interpretation, and (b) The
interpretation given in this paper.\newline
\newline
Fig. 3 The phase diagram for the vortex lattice structure as a function of
the four fold anisotropy parameter $\eta ^{\prime }$ and the two fold
anisotropy parameter $\lambda ^{\prime }$ .

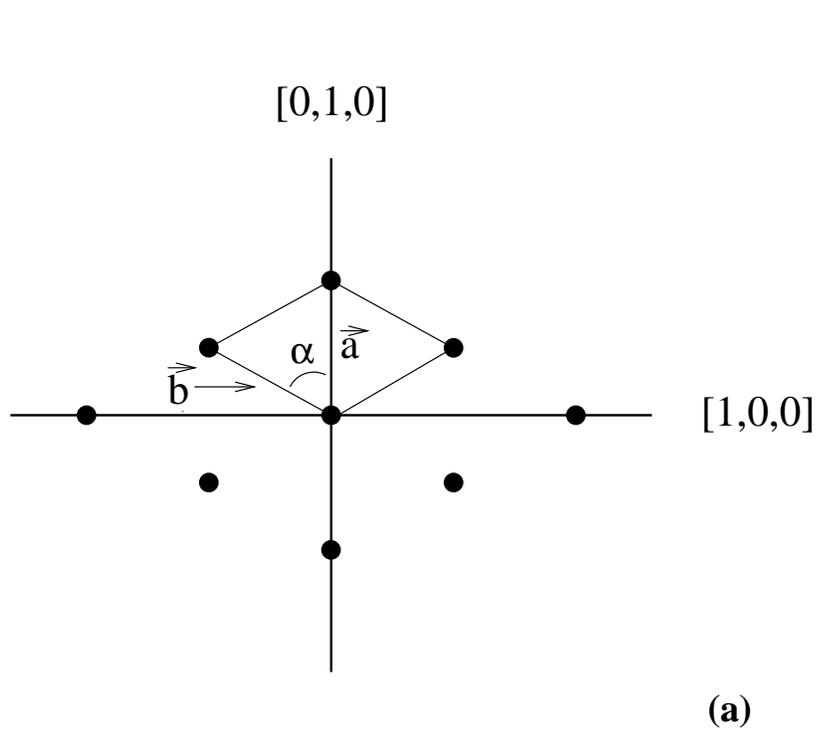
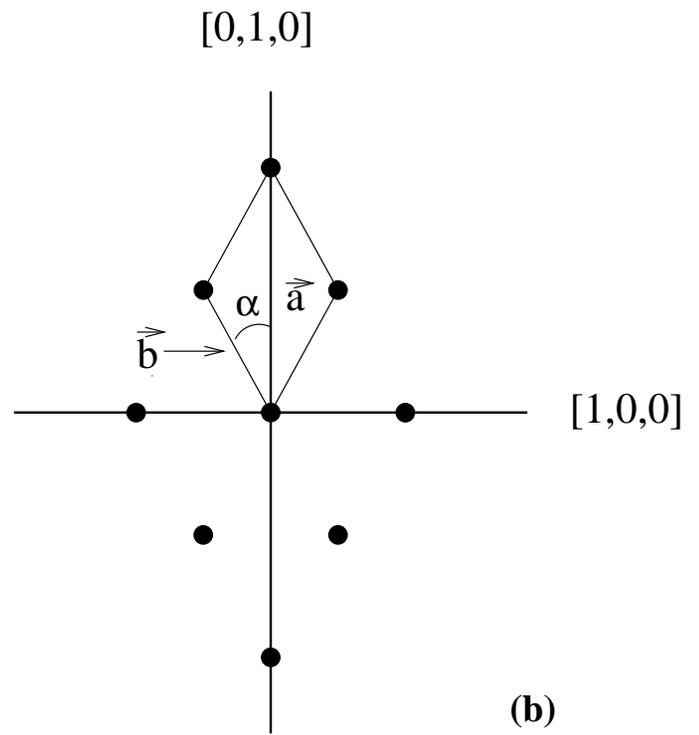
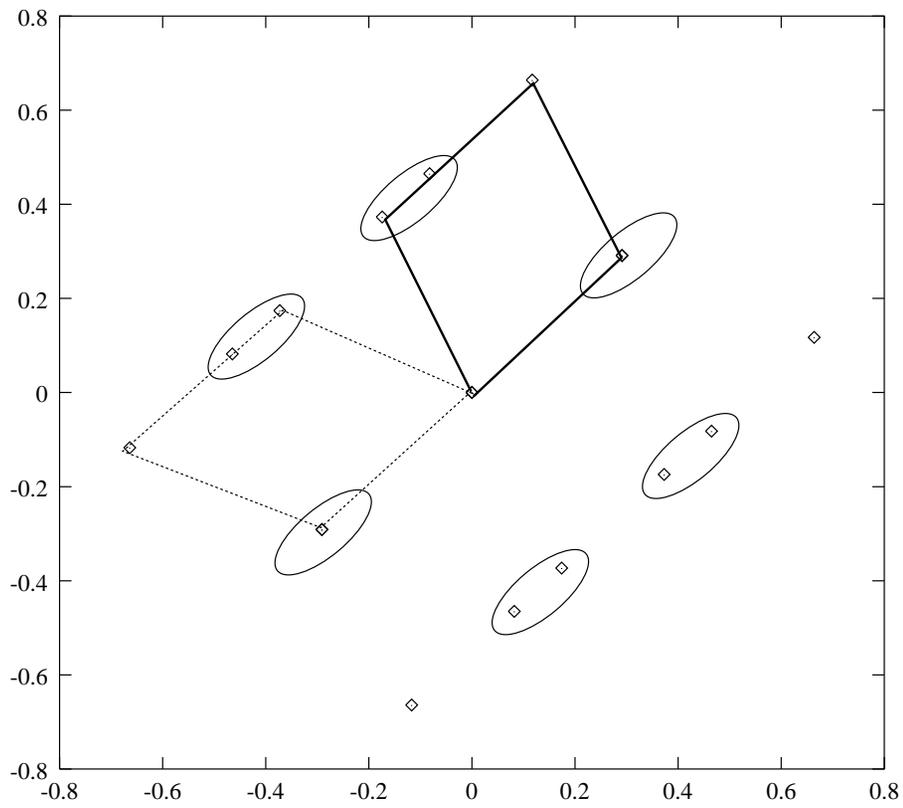
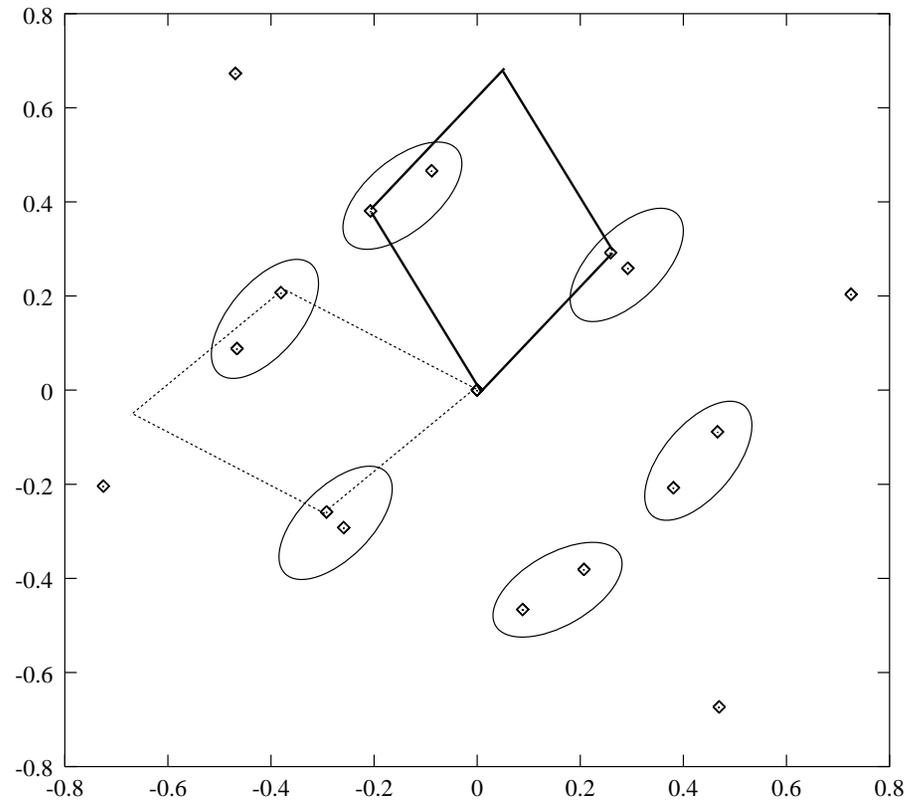
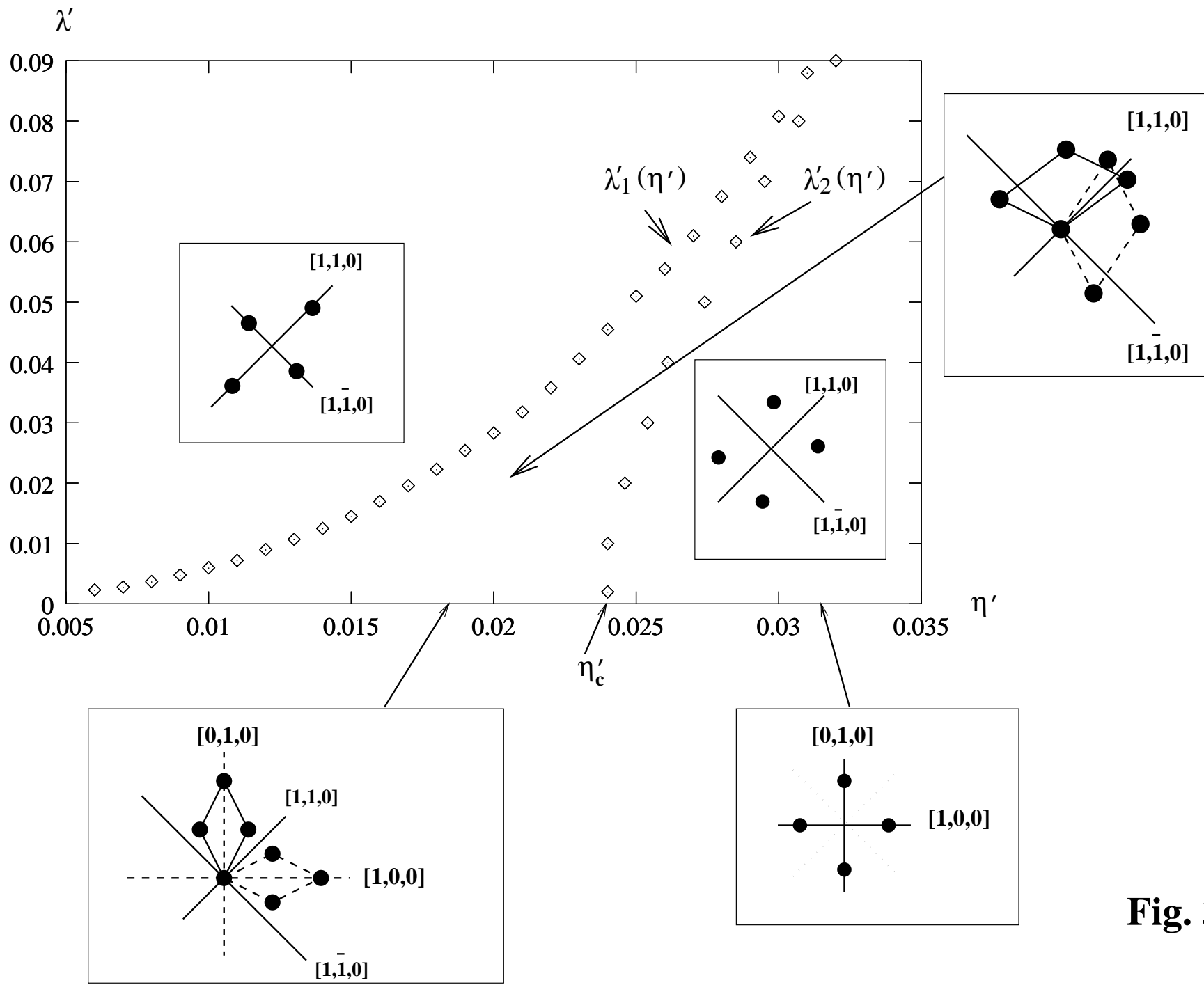
\end{document}